\documentstyle[aps]{revtex}
\begin{document}
\draft
\preprint{gr-qc/yymmxxx}
\title{Universal canonical black hole entropy}
\author{Ashok Chatterjee\footnote{email: ashok@theory.saha.ernet.in} 
and Parthasarathi Majumdar\footnote{On
deputation from the Institute of Mathematical Sciences, Chennai 600
113, India; email: partha@theory.saha.ernet.in}}
\address{Theory Group, Saha Institute of Nuclear Physics, Kolkata 700
064, India.}
\maketitle
\begin{abstract}
Non-rotating black holes in three and four dimensions are shown to
possess a canonical entropy obeying the Bekenstein-Hawking area law
together with a leading correction (for large horizon areas) given
by the logarithm of the area with a {\it universal} finite negative
coefficient, provided one assumes that the quantum black hole mass
spectrum has a power law relation with the quantum area spectrum
found in Non-perturbative Canonical Quantum General Relativity. The
thermal instability associated with asymptotically flat black holes
appears in the appropriate domain for the index characterising this
power law relation, where the canonical entropy (free energy) is
seen to turn complex.
\end{abstract}
\vglue .2in
The microcanonical entropy ($S_{MC}$) of generic stationary
non-rotating macroscopic four dimensional black holes, modelled as
{\it isolated} horizons \cite{aa1}, has been shown to obey, for large
fixed horizon areas, the Bekenstein-Hawking Area Law (BHAL)  
\cite{aa2} together with an infinite series of corrections, each term
of which is finite and calculable \cite{km}. The leading correction
is logarithmic in area (BH entropy) with a coefficient 
(calculated in \cite{km}) that has been
argued to be universal \cite{car}, \cite{gkv}. These results are
obtained within Non-perturbative Canonical Quantum General Relativity
(NCQGR), and crucially depend on the discrete spectrum of geometrical
observables, especially the area \cite{al}. 

It is well-known in elementary statistical mechanics \cite{hua}
that, unlike the canonical entropy, the definition of the
microcanonical entropy in fact is not unique. One can define it as
the logarithm of the {\it degeneracy} of quantum microstates
characterizing the system, as has been employed in the calculations
cited above \cite{aa2}, \cite{km}. Alternatively, one can define it
as the logarithm of the {\it density} of these microstates as a
function of the energy. While both these definitions yield the 
same area
law, they donot lead to the same logarithmic corrections. The
difference actually depends on the relation between the energy
(mass) of the black hole and the horizon area. This difference plays
an important role when the NCQGR result is used to compute the
effect of thermal fluctuations on the {\it canonical} entropy of
black holes \cite{cm},

In \cite{cm}, the precise distinction and relationship between 
corrections to the
BHAL arising due to quantum spacetime fluctuations, and due to
thermal fluctuations within a canonical framework (admittedly
heuristic), has been delineated. The logarithmic corrections due to
quantum spacetime fluctuations are interpreted as `finite size
effects'.\footnote{Computations of the entropy for matter fields in 
{\it fixed Euclidean} classical black hole backgrounds have been 
performed in 
\cite{fixed}. Our approach differs from these in that we focus on 
the {\it non-perturbative} quantum fluctuations of the 
{\it Lorentzian} black hole spacetime itself. This is also in 
contrast to the work of ref. \cite{byt} which is purely 
perturbative in nature.} Thermal fluctuations then generate 
additional log-area corrections which are computed in \cite{cm} for 
generic non-rotating
anti-de Sitter black holes. Fortunately, the thermal fluctuation
contributions to the canonical entropy can be expressed entirely in
terms of the microcanonical entropy which has been computed within
NCQGR as already mentioned. Unfortunately, however, the extra
logarithmic contribution to the canonical entropy owing to the
difference in the two definitions of the microcanonical entropy has
been missed out in \cite{cm}.

Note that within NCQGR the spectrum of the observable corresponding
to the classical isolated horizon mass is not known, while the
discrete spectrum of the area observable is completely understood
\cite{al}. Consequently, the difference between the two
microcanonical entropy definitions cannot actually be computed
within NCQGR. This also implies that the canonical ensemble cannot
be defined in the NCQGR formalism with as much rigour and
reliability as the microcanonical ensemble.

In this paper, we {\it postulate} that the two observables (mass 
and area) have a power-law relationship for large mass (area).
We then proceed {\it heuristically} to define a canonical partition
function in the standard manner. Restricting to macroscopic black
holes of large area (energy), this partition function is computed
using a saddle-point approximation and the canonical entropy
extracted from it. The difference in the two definitions of
microcanonical entropy now manifests as an extra logarithmic
correction to the BHAL for the canonical entropy, over and above
logarithmic terms arising from quantum spacetime fluctuations
\cite{km} and thermal fluctuations \cite{dmb}, \cite{cm}, \cite{gm}
(for adS black holes). One observes that, just like the earlier
microcanonical entropy result \cite{km}, the complete canonical
entropy also has a logarithmic correction to the BHAL with a
universal coefficient, {\it independent} of the index characterizing
the assumed power-law relation between the mass and the area.
However, the value of the index plays a crucial role in designating
the domain of validity of the assumption of thermal equilibrium, and
hence of the entire computation.

We should mention that for specific adS black holes, the extra
logarithmic correction to the canonical entropy has also been
suggested in \cite{gm} (as a kind of `Jacobian' factor), upon the
assumption of an equally-spaced area spectrum.\footnote{This
assumption, without the underpinning of a theory of quantum
spacetime fluctuations as here, is in fact adhoc.} In
addition, a power-law relation between the area and mass spectra has
been assumed, as here. The extra logarithmic correction to the
canonical entropy has been computed for specific adS black holes in
\cite{gm}, and the results appear to be in agreement with our
general formulation. The microcanonical entropy contribution has of
course not been considered at all in \cite{gm}. Thus, the actual
source of the extra log terms has not been traced in \cite{gm} to 
the difference in the two definitions of microcanonical entropy, as 
we have done here.  

The canonical partition function is given, as always, by
\begin{eqnarray}
Z_C(\beta)~=~Tr~\exp -\beta \hat{H} ~.
\label{qpf}
\end{eqnarray}
where, as argued in section IV of \cite{cm}, only the part of the 
Hamiltonian pertaining to the (inner) boundary of spacetimeis 
relevant for 
thermodynamics, the bulk part being rendered redundant by the 
quantum Hamiltonian constraint. Taken together with caveats mentioned 
above, one can heuristically reexpress eq. (\ref{qpf}) as a sum
\begin{eqnarray}
Z_C~=~\sum_p g(E(A_p))~\exp -\beta E(A_p)~, \label{suma}
\end{eqnarray}
where, $A_p$ is the allowed area of a very large black hole horizon 
with $p~,~(p \gg 1)$ punctures due to the impinging of spin network 
edges on the 
horizon; each puncture carries a spin 1/2. The area spectrum, as 
given by the NCQGR formalism then implies that $A_p\sim p$. We 
further assume that the energy (mass) of the black hole is a {\it 
monotonic} function of the area eigenvalue. $g(E(A_p))$ is the 
degeneracy of the energy level designated by $E(A_p)$. It is 
important to keep in mind that eq. (\ref{suma}) is actually valid 
only for very large (average) areas, i.e., for large 
(average) energies, since only in the limit of large mass or areas 
do black 
holes ever have a chance of being in thermal equilibrium with 
radiation. Thus the sum (\ref{suma}) is dominated by large values of 
$p \gg 1$. Using now the Poisson resummation formula
\begin{eqnarray}
\sum_{n=-\infty}^{\infty} f(n)~=~\sum_{m=-\infty}^{\infty} 
\int_{-\infty}^{\infty} dx ~\exp (-2\pi i mx)~f(x)~, \label{poi}
\end{eqnarray}
one obtains
\begin{eqnarray}
Z_C~\simeq~\int_{-\infty}^{\infty} dx~g(E(A(x)))~\exp -\beta 
E(A(x))~, \label{appr}
\end{eqnarray}
where, we have set $\sum_l \exp - 2\pi i lx \approx 1$ for $|x| \gg 
1$, as appropriate to very large black holes. Changing integration 
variables from $x$ to $E$ we obtain,
\begin{eqnarray}
Z_C~&=&~\int dE~\left({dE \over dx} \right)^{-1}~g(E)~\exp -\beta E 
~\nonumber \\
~&=&~\int dE~\exp [S_{MC}(E)~-~\log|{dE \over dx}|~-~\beta E]~ .
\label{newp}
\end{eqnarray}
In eq. (\ref{newp}), the microcanonical entropy $S_{MC} \equiv \log 
g(E)$ as used in \cite{aa2} and \cite{km}to calculate the entropy of 
non-rotating isolated horizons; this is {\it not} the 
definition used in \cite{cm} where instead we identified the 
logarithm of the density of states as the microcanonical entropy. In 
doing so, the second term in the exponent above, namely $-\log |{dE 
\over dx}|$ has been missed in \cite{cm}. In other words, 
defining ${\tilde S}_{MC} \equiv \log \rho(E)$, the two definitions 
of microcanonical entropy are related by 
\begin{eqnarray}
{\tilde S}_{MC}(E)~=~S_{MC}(E)~-~\log|{dE \over dx}|~.
\end{eqnarray}
Obviously, the extra contribution is irrelevant for demonstrating 
the BHAL, but since we are interested in ascertaining log-area 
corrections to the BHAL for canonical entropy, the effect of the 
second term on the {\it rhs} above must be investigated. 

We proceed as in \cite{cm} by evaluating the integral in a 
saddle-point approximation to yield, for the saddle-point $E=M$
\begin{eqnarray}
Z_C~\simeq~ \exp \left\{ S_{MC}(M) - \beta M - \log|{dE \over 
dx}|_{E=M}] \right \}~\left [{\pi \over -S_{MC}''(M)} \right]^{1/2}  
~. \label{sad}
\end{eqnarray}
The canonical entropy is now straightforward to extract, using the 
standard thermodynamic formula $S_C = \log Z_C + \beta M$, leading to
\begin{eqnarray}
S_C~&=&~S_{MC}(M)~-~\log|{dE \over dx}|_{E=M}~-~\frac12 \log 
[-S_{MC}''(M)]~+~const.~\nonumber \\
~&=&~S_{MC}(M)~-~\frac12 \log(-\Delta)~, \label{cane} 
\end{eqnarray}
where,
\begin{eqnarray}
\Delta~\equiv~{d^2 S_{MC} \over dE^2}~\left({dE \over 
dx}\right)^2|_{E=M}~ . 
\label{delta}
\end{eqnarray}
Manipulating the derivatives in eq. (\ref{delta}) using the chain 
rule several times, it now follows that
\begin{eqnarray}
\Delta~=~\left[ {d^2 S_{MC} \over dA^2}~-~\left({dS_{MC} \over 
dA}\right) {d^2 E/dA^2 \over dE/dA} 
\right]~\left({dA \over dx}\right)^2|_{E=M}~. \label{dell}
\end{eqnarray}

Since $S_{MC}$, computed within NCQGR as a function of the
horizon area (\cite{aa2}, \cite{km}) contains universal
logarithmic corrections \cite{km}, \cite{car}, and the area spectrum
for large isolated horizons is known to be linear as a 
function of $x$ \cite{al}, essentially the only `nonuniversal' aspect 
of the canonical entropy above is the second term within the square 
brackets which involves the energy (mass)-area relation. We now 
hypothsize that this relation is a power law one (for large area),
\begin{eqnarray}
E(A)~=~const.~A^r~, \label{ena}
\end{eqnarray}
where, $r$ is a real number, positive or negative. This relation, 
together with the formula for microcanonical entropy \cite{km}
\begin{eqnarray}
S_{MC}~=~S_{BH}~-~\frac32 \log S_{BH}~+~const.~+~O(S_{BH}^{-1})~, 
\label{mics}
\end{eqnarray}
yields the canonical entropy given by the BHAL and {\it universal} 
logarithmic corrections,
\begin{eqnarray}
S_C~=~S_{BH}~-~\log S_{BH}~-~\frac12 \log(r-1)~+~const.~+~O(A^{-1})~. 
\label{fcane}
\end{eqnarray}

It is of interest to note that the leading logarithmic correction is
actually independent of the exponent $r$ characterizing the
mass-area relation, which is different for different classes of
black holes.  However, this exponent plays a most crucial role: it
decides whether the black hole is in thermal equilibrium or not.
E.g., for asymptotically flat non-rotating black holes of large
macroscopic horizon size, $r=1/2 < 1$. This implies that the
canonical entropy (and the Gibbs free energy) has turned {\it
complex} for such black holes, clearly signifying a thermal
instability. The saddle-point approximation used to extract the
canonical entropy breaks down whenever $r < 1$.

On the other hand, {\it all} generic non-rotating anti-de Sitter
black holes have $r > 1$ in the parameter domain characterized by a
large fixed cosmological constant. This is basically the domain in
which the horizon radius is far larger than the length parameter
$\ell \sim \Lambda^{-1/2}$.  Thus, for the non-rotating BTZ black
hole, $r=2$ while for the adS-Schwarzschild black hole in four
dimensions, $r=3/2$, so that thermal equilibrium can indeed ensue in
this domain of parameter space. Incorporation of electric charge
seems to involve no extra complication generically. As the exponent
$r \rightarrow 1$ from above, for the adS black holes one appears to
approach the Hawking-Page phase transition \cite{hawp} from the
black hole `phase' to an adS `gas phase'. 

Consider now an alternative approach to the canonical entropy, based
not on a canonical energy ensemble, but rather a canonical {\it area}
ensemble \cite{kras}. This is motivated by the observation already
made, viz., the mass of the horizon is not a well-defined observable
in NCQGR, while the area is. It is for this reason that the
microcanonical entropy within this formalism automatically yields a
result as a function of the horizon area, without any reference to
the spectrum of any energy operator. To define a canonical partition 
function
in such an area ensemble, we shall have to require that the total
area (just like the total energy in the standard canonical ensemble)  
is a constant of motion. To achieve this, we may think of area 
fluctuations being caused by the addition or deletion of deficit 
angles made on the horizon 2-sphere by the appearance of new spin 
network edges impinging on the horizon, or old edges disappearing, by 
a yet unknown mechanism. All these fluctuations keep a `total' area 
fixed.

The partition function for such an area ensemble is given by 
\cite{kras}
\begin{eqnarray}
Z(\alpha)~&=&~Tr \exp -\alpha {\hat A} ~\nonumber \\
&=&~\sum_p g(A_p)~\exp -\alpha A_p~,
\label{apf}
\end{eqnarray}
where $\alpha$ is the corresponding inverse ambient 
`temperature'.\footnote{Brown et. al. \cite{bro} have referred to a 
quantity analogous to $\alpha$, i.e., the canonical conjugate to 
area, as the `surface pressure'.} The relation of $\alpha$ to the 
inverse temperature $\beta$ in a standard canonical ensemble is far 
from clear. $p$ is the number of punctures; we assume 
as before that in the formula (\ref{apf}) above, the {\it rhs} is 
dominated by terms of large $A_p$, assumed to be 
linearly dependent on large $p$. Appealing to the Poisson 
resummation formula as before, the partition function can be 
rewritten as an 
integral over area
\begin{eqnarray}
Z(\alpha)~=~\int dA ~\exp \left \{S_{MC}(A)~-~\log |{dA \over dx}| 
~-~\alpha A \right \}~. \label{aint}
\end{eqnarray}
The integral can be evaluated as before by a saddle point 
approximation, yielding,
\begin{eqnarray}
Z(\alpha)~=~\exp \left \{ S_{MC}(A_0)~-~\log |{dA \over dx}|_{A=A_0} 
~-\alpha A_0 \right \}~\left[{\pi \over -S_{MC}''(A_0)} 
\right]^{1/2}~,
\label{sada}
\end{eqnarray}
where $A_0$ is the saddle point (equilibrium) area. Observe at this 
point that, given that the area spectrum is 
linear in $x$ for large $x$, the $|dA/dx|^{-1}$ term in the integral 
is a constant independent of the area. This is quite unlike the 
earlier situation in the standard canonical ensemble where this 
particular Jacobian term had nontrivial implications for the 
canonical entropy. Furthermore, the last factor in eq, (\ref{sada}), 
representing the contribution due to thermal fluctuations, is once 
again completely determined by the microcanonical entropy calculated 
within NCQGR.  

However, observe that using eq. (\ref{mics}) one immediately gets
$S_{MC}''(A_0) > 0$. It follows that within the canonical area 
ensemble,
it is not possible to use the saddle-point approximation to
calculate the effect of Gaussian thermal fluctuations on logarithmic
corrections to the canonical entropy of {\it all} general
relativistic black holes. The thermal instability discerned in the
standard `energy' canonical ensemble for asymptotically flat black
holes now appears universally for this area ensemble, This raises
the question as to whether the area ensemble is at all an
equilibrium thermodynamic ensemble for black holes of large area.

In conclusion, we reiterate that the universality of the logarithmic
correction found by us in the standard canonical ensemble, emerges
{\it without} any information about the actual mass spectrum of the
black hole as a function of the area. It is enough to assume the
mere existence of a power law relation. Furthermore, no additional
assumption at all needs to be made regarding the nature of the area
spectrum which is completely determined within the NCQGR formalism
\cite{al}. The fact that the exponent characterizing the power law
area-mass relation is decisive in delineating the domain of thermal
equilibrium vis-a-vis black holes is also an interesting, albeit an
expected, feature. It is however, extremely important to keep in
mind the caveat in all of the above, of extending a classical
relation between mass and area of black holes to the spectra of the
corresponding quantum operators. Elimination or otherwise of this 
caveat remains the top priority issue in future work on this 
problem.

\end{document}